\documentclass{aastex63}
\usepackage{epstopdf}
\usepackage{graphicx}
\usepackage{multirow}

\shorttitle{Formation and Evolution of Transient Bubbles Driven by Erupting Mini-filaments}
\shortauthors{Guo et al.}

\begin{document}

\title{Formation and Evolution of Transient Prominence Bubbles Driven by Erupting Mini-filaments}

\correspondingauthor{Yijun Hou}
\email{yijunhou@nao.cas.cn}

\author[0000-0001-9893-1281]{Yilin Guo}
\affiliation{Beijing Planetarium, Beijing Academy of Science and Technology, Beijing 100044, China}

\author[0000-0002-9534-1638]{Yijun Hou}
\affiliation{National Astronomical Observatories, Chinese Academy of Sciences, Beijing 100101, China}
\affiliation{School of Astronomy and Space Science, University of Chinese Academy of Sciences, Beijing 100049, China}
\affiliation{Yunnan Key Laboratory of the Solar physics and Space Science, Kunming 650216, China}
\affiliation{State Key Laboratory of Solar Activity and Space Weather, Beijing 100190, China}

\author[0000-0001-6655-1743]{Ting Li}
\affiliation{National Astronomical Observatories, Chinese Academy of Sciences, Beijing 100101, China}
\affiliation{School of Astronomy and Space Science, University of Chinese Academy of Sciences, Beijing 100049, China}
\affiliation{State Key Laboratory of Solar Activity and Space Weather, Beijing 100190, China}

\author[0000-0001-9493-4418]{Yuandeng Shen}
\affiliation{Yunnan Observatories, Chinese Academy of Sciences, Kunming Yunnan 650216, China}
\affiliation{Yunnan Key Laboratory of the Solar physics and Space Science, Kunming 650216, China}

\author{Jincheng Wang}
\affiliation{Yunnan Observatories, Chinese Academy of Sciences, Kunming Yunnan 650216, China}
\affiliation{Yunnan Key Laboratory of the Solar physics and Space Science, Kunming 650216, China}

\author{Jun Zhang}
\affiliation{School of Physics and Optoelectronics Engineering, Anhui University, Hefei 230601, China}

\author{Jianchuan Zheng}
\affiliation{Shenzhen Astronomical Observatory, Shenzhen 518040, China}

\author{Dong Wang}
\affiliation{Shenzhen Astronomical Observatory, Shenzhen 518040, China}

\author{Lin Mei}
\affiliation{Shenzhen Astronomical Observatory, Shenzhen 518040, China}

\begin{abstract}
Prominence bubbles, the dark arch-shaped ``voids'' below quiescent prominences, are generally believed to be caused by the interaction between the prominences and the slowly-emerging or quasi-stable underlying magnetic loops. However, this scenario could not explain some short-lived bubbles with extremely dynamic properties of evolution. Based on high-resolution H$\alpha$ observations, here we propose that bubbles should be classified into two categories according to their dynamic properties: quasi-steady (Type-I) bubbles and transient (Type-II) bubbles. Type-I bubbles could remain relatively stable and last for several hours, indicating the existence of a quasi-stable magnetic topology, while Type-II bubbles grow and collapse quickly within one hour without stability duration, which are usually associated with erupting mini-filaments. Analysis of several typical Type-II bubbles from different views, especially including an on-disk event, reveals that Type-II bubbles quickly appear and expand at a velocity of $\thicksim$5--25 km s$^{-1}$ accompanied by an erupting mini-filament below. The mini-filament's rising velocity is slightly larger than that of the Type-II bubbles' boundary, which will lead to the collision with each other in a short time, subsequent collapse of Type-II bubbles, and formation of a large plume into the above prominence. We also speculate that only if the angle between the axis of the erupting mini-filament and the line-of-sight is large enough, the interaction between the erupting mini-filament and the overlying prominence could trigger a Type-II bubble with a typical arch-shaped but quickly-expanding bright boundary.
\end{abstract}


\section{Introduction}

Solar prominences are a common but intriguing phenomenon in the solar atmosphere. They consist of cool and dense plasma and suspend in hot and tenuous corona above magnetic neutral lines. When observed above the solar limb, prominences appear as emission features in chromospheric lines, whereas they are identifiable as dark features on the solar disk and are known as ``filaments'' \citep{2010SSRv..151..333M,2014LRSP...11....1P,2016ApJ...823...22X,2020RAA....20..166C,2020A&A...640A.101H,2021ApJ...920...77G}. According to their formation location, the prominences are divided into three types \citep{2014LRSP...11....1P,2015ASSL..415.....V}: active-region prominences nearby sunspots \citep{2012A&A...539A.131K,2015ApJS..219...17Y,2018A&A...619A.100H}, intermediate prominences around the borders of active region \citep{2014ApJ...786L..16J}, and quiescent prominences in quiet regions of the Sun \citep{2016A&A...589A.114L,2022ApJ...934L...9L}.

Recent high-resolution observations reveal two fascinating phenomena in quiescent prominences: prominence bubbles and plumes \citep{2008ApJ...676L..89B,2010ApJ...716.1288B}. Prominence bubbles are dark arch-shaped ``voids'' under the prominences observed in strong chromospheric lines. Bright arch-like boundaries are typical features between the bubbles and the overlying prominences. Small-scale upward plumes usually originate from the bubble boundaries. Bubbles and plumes are suggested as \textbf{a} probable way of transport of magnetic flux and helicity from chromospheric layer into the overlying prominence \citep{2011Natur.472..197B}, which could eventually lead to the eruption of prominence and launching coronal mass ejection \citep{2006ApJ...644..575Z}. Therefore, studying the formation and evolution of bubbles and plumes is helpful for revealing the magnetic topology and dynamic evolution of prominences.

Formation of bubbles is generally believed to be connected with emerging magnetic flux under the prominences. \citet{2011Natur.472..197B} speculated that bubbles originate from the emergence of twisted magnetic flux below prominences. Based on numerical simulations, \citet{2012ApJ...761....9D} and \citet{2014AA...567A.123G} created the magnetic topology of bubble with arcade field lines of a parasitic bipole inserted under a linear force-free magnetic flux rope. \citet{2015ApJ...814L..17S} proposed a similar carton model that piled-up magnetic dips supported by the low-lying closed magnetic system form the bubble structure. Recently, combining direct on-disk photospheric magnetic field observations, \citet{2021ApJ...911L...9G} reconstructed 3D magnetic topology of on-disk bubbles through nonlinear force-free field extrapolations, and suggested that the bubble boundaries correspond to interfaces between prominence magnetic dips and magnetic loops rooted nearby. Although the existing studies have proved that magnetic loops interacting with the upper prominence could support the stable existence of bubbles, due to lack of enough photospheric magnetic field observations of on-disk bubbles and data of bubble formation process, several open questions about bubbles still remain: What is the formation mechanism of bubbles? Do all bubbles have the same magnetic nature and formation mechanism? If not, then what are the key factors determining the formation of these different types of bubbles and their evolution characteristics?

Existing studies about bubbles pursue a universal model that can perfectly explain the physical nature of observational features of all prominence bubbles. However, by revisiting previous studies and examining a large number of high-resolution prominence observations from the New Vacuum Solar Telescope \citep[NVST;][]{2014RAA....14..705L,2020ScChE..63.1656Y}, we find that prominence bubbles in different events display obvious diversity of dynamic behavior. As reported by most researchers \citep{2010ApJ...716.1288B,2011Natur.472..197B,2017ApJ...850...60B,2012ApJ...761....9D,2012ASPC..456...77S,2014AA...567A.123G,2015ApJ...814L..17S,2021ApJ...923L..10C,2022AA...659A..76W,2021RAA....21..222X}, the rising velocity of bubbles is no more than 3 km s$^{-1}$ over half an hour or several hours of evolution, indicating that bubbles can keep almost stable or evolve very slowly. On the contrary, there are some bubbles with faster rising velocity and shorter lifetime. For example, \citet{2008ApJ...687L.123D} reported that some bubbles with a rising bright core below can reach velocity of up to $\thicksim$12--13 km s$^{-1}$ and disappear within 30--50 minutes. As a result, we could conclude that some bubbles could remain relatively stable during a long period of time, but others own extremely dynamic properties of evolution.

Although existing model that the slowly-emerging magnetic loops interacting with the upper prominence dips form a quasi-stable structure supports an existence of stable bubbles well, this scenario could not explain formation mechanism and evolution process of extremely dynamic bubbles. Therefore, in this paper, based on observations from the NVST and Shenzhen Astronomical Observatory (SZAO), according to the obvious diversity of dynamic behavior, we firstly classify prominence bubbles into two categories: quasi-steady (Type-I) bubbles and transient (Type-II) bubbles, and then compare and analyze the formation mechanisms and dynamic evolution characteristics of the two types of bubbles, respectively.

\section{Observations}

Based on NVST and SZAO observations of prominence bubbles, we investigate two types of bubble events showing different evolution characteristics. Five bubble events are studied in detail, including one Type-I bubble and four Type-II bubbles. The NVST is a 1 m vacuum solar telescope at the Fuxian Lake Solar Observatory (China). It has a multi-channel high resolution imaging system, including a channel for chromosphere (H$\alpha$ 6562.8 {\AA}) and two channels for photosphere (TiO 7058 {\AA} and \emph{G}-band 4300 {\AA}). The H$\alpha$ filter is a tunable Lyot filter with a bandwidth of 0.25 {\AA}. In this paper, we employ the NVST H$\alpha$ data, which are processed through dark current subtraction, flat field correction, and speckle masking \citep{1977OptCo..21...55W,1983ApOpt..22.4028L,2016NewA...49....8X}. Additionally, we also analyze the H$\alpha$ observations from an 18 cm aperture refractor telescope on SZAO. The focal length of the telescope is 1.26 m, and the actual focal length is 5.04 m after adding four times the Barlow lens. The telescope equips an Energy-Rejection-Filter at the objective end of telescope, a Daystar Quantum PE (H$\alpha$ 6562.8 {\AA}; FWHM: 0.3 {\AA}) at the rear end, and a 1608$\times$1140 pixel CCD of Apollo-M MAX. Prominence images are firstly observed with video mode (200 frames per video in ser format), and then are reconstructed by lucky imaging to stack 150 frames with AutoStakkert software and sharping the frames with RegiStax software. Detailed information on the related observations is summarized in Table \ref{t1}.

In addition, a 193 {\AA} image from the Atmosphere Imaging Assembly \citep[AIA;][]{2012SoPh..275...17L} on board the Solar Dynamics Observatory \citep[SDO;][]{2012SoPh..275....3P} and an H$\alpha$ image from Chinese H$\alpha$ Solar Explorer \citep[CHASE;][]{2022SCPMA..6589602L} were used to display the location of bubbles on the solar limb. Their spatial resolutions are 0.{\arcsec}6 pixel$^{-1}$ and 1.{\arcsec}04 pixel$^{-1}$, respectively.

\begin{table*}
\caption{Data Summary}
\centering
\begin{tabular}{clcccccc}
\hline\hline
\textbf{Bubble} & \textbf{Date} & \textbf{Time (UT)} & \textbf{Telescope} & \textbf{Position}& \textbf{Spatial Sampling} & \textbf{Cadence}&\textbf{Figures}\\
\hline\hline
Type-I&10-Nov-2018&05:10--06:20&NVST&51{\degr}S, 80{\degr}W&0.{\arcsec}136 pixel$^{-1}$&30 s&\ref{fig1}\\
\hline
\multirow{4}*{Type-II}&9-Mar-2023&06:40--08:20&SZAO&40{\degr}S, ---{\degr}E&0.{\arcsec}3 pixel$^{-1}$&60 s&\ref{fig1}, \ref{fig2}\\
\cline{2-8}
~&9-Mar-2023&08:50--09:20&SZAO&40{\degr}S, ---{\degr}E&0.{\arcsec}3 pixel$^{-1}$&60 s&\ref{fig3}\\
\cline{2-8}
~&19-May-2019&03:59--05:01&NVST&35{\degr}S, 88{\degr}E&0.{\arcsec}165 pixel$^{-1}$&41 s&\ref{fig3}\\
\cline{2-8}
~&7-Nov-2018&03:25--04:10&NVST&35{\degr}N, 26{\degr}E&0.{\arcsec}136 pixel$^{-1}$&26 s&\ref{fig3}\\
\hline
\label{t1}
\end{tabular}

\tablecomments{
The New Vacuum Solar Telescope (NVST) is a 1 m vacuum solar telescope at the Fuxian Lake Solar Observatory (China). The Shenzhen Astronomical Observatory (SZAO; China) owns an 18 cm aperture refractor telescope on it. The NVST and SZAO H$\alpha$ data is used in this paper.}

\end{table*}

\section{Results}

\subsection{Classification of Bubbles: Quasi-steady Bubbles and Transient Bubbles}

Based on high-resolution H$\alpha$ observations from NVST and SZAO, we find several especial bubbles whose dynamic evolution is rather different from bubbles reported before. Therefore, according to different physical properties and dynamic evolutions of bubbles, we give a new perspective that prominence bubbles can be classified into two categories: quasi-steady (Type-I) bubbles and transient (Type-II) bubbles. The Type-I bubbles with relatively stable boundaries can last for half an hour or even several hours and have been extensively studied through high-resolution observations, e.g., observations from Hinode satellite and NVST \citep{2010ApJ...716.1288B,2011Natur.472..197B,
2012ASPC..454...79B,2017ApJ...850...60B,2010SoPh..267...75R,
2012ApJ...761....9D,2012ASPC..456...77S,2014AA...567A.123G,
2015ApJ...814L..17S,2021ApJ...923L..10C,2022AA...659A..76W,
2021RAA....21..222X}. As for the Type-II bubbles, they can rise quickly within one hour and have been rarely reported. The only two possible Type-II bubble events in the literature were reported by \citet{2008ApJ...687L.123D}, where bubbles were driven by a bright core, which might be compressed plasma trapped in closed field, and rose quickly with a velocity of $\thicksim$12--13 km s$^{-1}$. However, in that study, these short-lived bubbles with quick-expanding boundaries were described and studied as typical Type-I bubbles. As a result, we claim that the concept of transient (Type-II) bubbles is proposed for the first time in the present study.

\begin{figure}[ht!]
\centering
\includegraphics [width=0.95\textwidth]{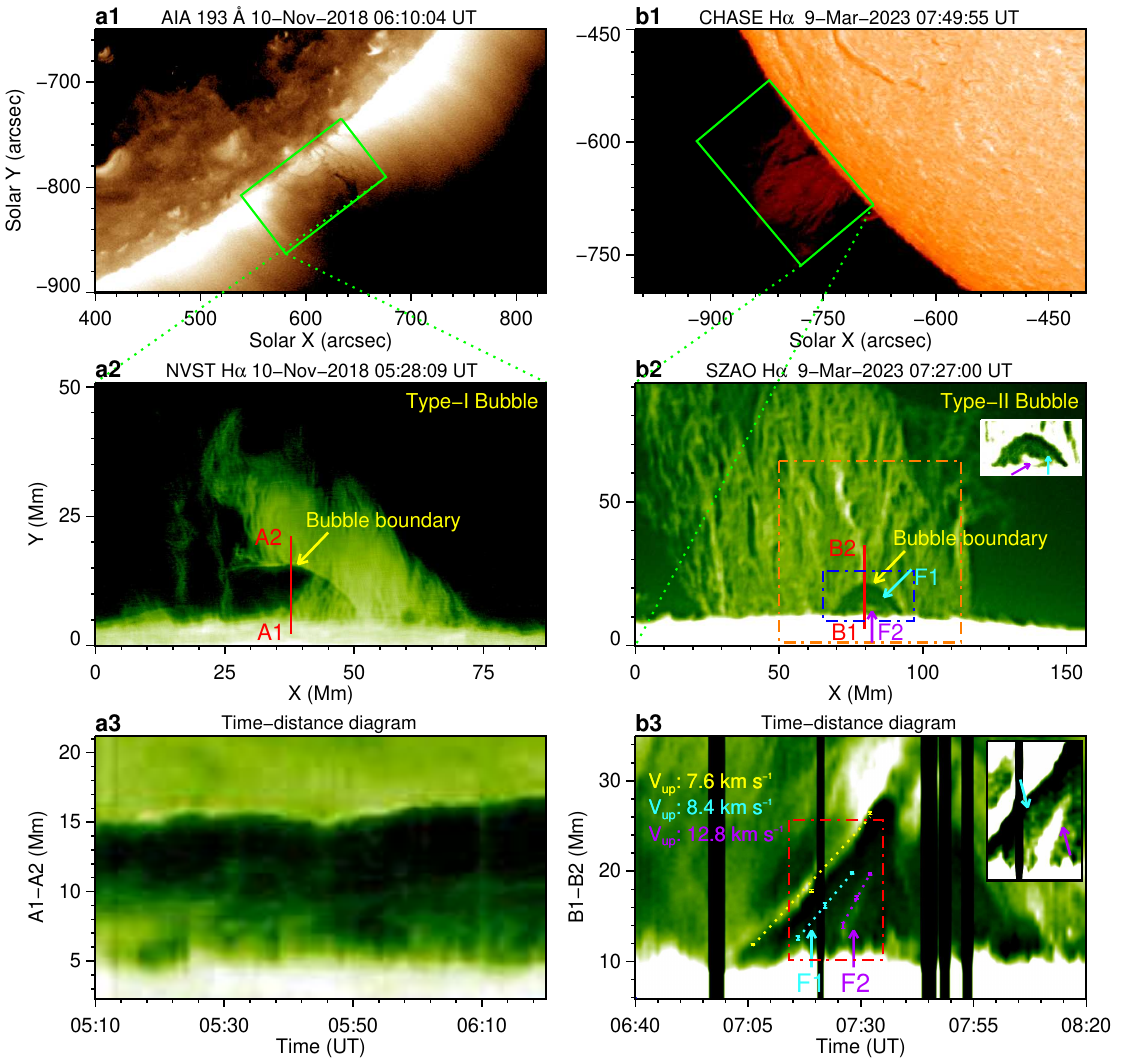}
\caption{Overview of a typical Type-I bubble (left panels) and Type-II bubble (right panels). (a1)\textbf{--}(a2) SDO/AIA 193 {\AA} and NVST H$\alpha$ images displaying the Type-I bubble. (a3) Time-distance plot derived from NVST H$\alpha$ images along the cut ``A1--A2''. (b1)\textbf{--}(b2) CHASE H$\alpha$ and SZAO H$\alpha$ observations showing the Type-II bubble. (b3) Time-distance plot derived from SZAO H$\alpha$ images along the cut ``B1--B2''. The green rectangles in panels (a1) and (b1) outline the field of view (FOV) of panels (a2) and (b2). The yellow arrows in panels (a2) and (b2) mark the boundaries of two bubbles. The orange dashed square in panel (b2) outlines the FOV of Figure 2. The images in the blue and red dashed boxes are enhanced contrast to display in the inset diagrams in the top right of panels (b2)--(b3). The cyan and purple arrows in the panels (b2)--(b3) mark the mini-filaments F1 and F2. The vertical black strips are gaps in observations. An animation of NVST and SZAO H$\alpha$ images is available online.
}
\label{fig1}
\end{figure}

\subsection{Comparison of Typical Quasi-steady Bubbles and Transient Bubbles}

In Figure 1, typical examples of Type-I and Type-II bubble are displayed. The Type-I bubble was observed at the southwest solar limb on 10 November 2018 (Figure 1(a1)). As presented in high-resolution NVST H$\alpha$ image with a smaller field of view (FOV), this bubble appears as a semi-circular void with a sharp arch-like boundary (marked by the yellow arrow in Figure 1(a2)) under a quiescent hedgerow prominence. To study the kinematic characteristics of the bubble, we derive a time-distance plot from sequences of NVST H$\alpha$ images along the cut``A1\textbf{--}A2'' (Figure 1(a3)). It is displayed that the bubble boundary kept almost stable from 05:10 UT to 06:20 UT. In addition, bubbles with similar quasi-steady evolution property were also reported by \citet{2015ApJ...814L..17S} and \citet{2017ApJ...850...60B}. Significantly, because NVST H$\alpha$ images do not cover the complete evolution process of bubble from generation to collapse, we estimate the lifetime of Type-I bubble by checking the observations from Global Oscillation Network Group \citep[GONG;][]{1996Sci...272.1284H} and SDO. The lifetime of Type-I bubble is summarized in Table \ref{t2}.

The Type-II bubble occurred below a quiescent hedgerow prominence at the southeast solar limb on 9 March 2023 (Figure 1(b1)). The SZAO H$\alpha$ image exhibits the transient bubble with a smaller FOV (Figure 1(b2)). This bubble has a sharp bright boundary as well. However, different from the Type-I bubble, this Type-II bubble appeared, grew, and collapsed quickly within $\thicksim$40 minutes, under which are two erupting mini-filaments (F1 and F2, marked by the cyan and purple arrows, respectively). The time-distance plot drawn from sequences of SZAO H$\alpha$ images along the cut ``B1--B2'' in Figure 1(b3) indicates that the Type-II bubble generated and rose into the prominence quickly with a velocity in plane of sky (POS) of 7.6 km s$^{-1}$. It grew up to about 17 Mm in less than 35 minutes. The POS velocity of F1 and F2 are 8.4 km s$^{-1}$ and 12.8 km s$^{-1}$, respectively, which are slightly higher than that of the bubble boundary.

\begin{figure}[ht!]
\centering
\includegraphics [width=0.8\textwidth]{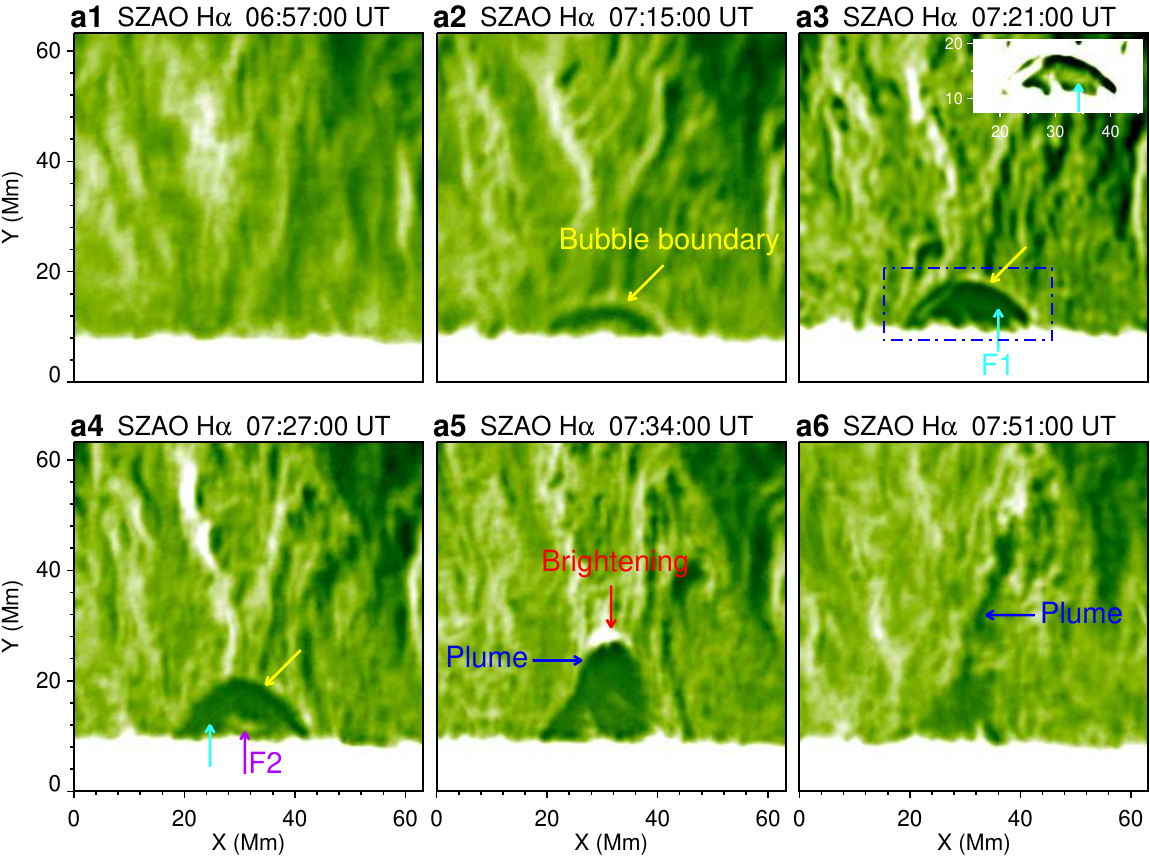}
\caption{Formation and collapse of the Type-II bubble driven by erupting mini-filaments F1 and F2. The arrows mark the bubble boundary  (yellow), mini-filaments F1 (cyan) and F2 (purple), brightening at the bubble boundary (red), and plume (blue) launched at the bubble boundary. The image in the blue box are enhanced contrast to display in the inset diagram in the top right of panel (a3).
}
\label{fig2}
\end{figure}

Comparing the two types of bubbles, we can find that their morphological characteristics are similar: they are both located below a quiescent prominence and have a bright arch-like boundary. However, the evolution properties of these two types of bubbles are vastly different: the Type-I bubble keeps relatively stable for more than one hour, while the Type-II grows quickly from its formation to the largest scale, and eventually collapses. Such difference could be caused by their different formation mechanisms and magnetic nature. The generally accepted magnetic topology model that bubbles are caused by the interaction between the prominences and underlying slowly-emerging or quasi-stable magnetic loops supports the stable existence of the Type-I bubbles reasonably. Different from the Type-I bubbles, the Type-II bubbles seem to have a close relation with the erupting mini-filaments below. The physical nature of Type-II bubbles will be further explored in the next subsection.

\subsection{Rapid Evolution of Transient Bubbles Driven by Mini-filament Eruption}

Based on SZAO H$\alpha$ observations, the formation and collapse processes of the Type-II bubble are analyzed and displayed in detail in Figure 2. At 06:57 UT, there was no bubble under the prominence (Figure 2(a1)). During the following several minutes, a bubble with a bright arch-like boundary was formed at the base of prominence (Figure 2(a2)). Meanwhile, below the bubble boundary, a mini-filament F1 appeared on the solar limb, which was clearly exhibited in Figure 2(a3). Soon after, another mini-filament F2 emerged below F1 with a higher velocity (Figure 2(a4)). As the F1 and F2 continually rose, it was obvious that the bubble expanded quickly from a flat void into a semicircle with a maximum height of 17 Mm. At around 07:33 UT, the erupting F1 and F2 merged with the bubble boundary. This merging likely caused the formation of a big plume and the collapse of the bubble. Figure 2(a5) displays that the plume began as a bulge at the top of the bubble boundary with an obvious brightening. Then, the plume rose, became more turbulent with finger-like structures, and faded into the prominence. However, the brightening was not observed in SDO/AIA channels, which suggests that there are no strong magnetic reconnections occurring between the mini-filaments and overlying prominence. Therefore, we cannot give a definite conclusion that the brightening is caused by a weak magnetic reconnection or plasma density enhancement, just like the uncertainty of formation mechanism of bright bubble boundary \citep{2012ApJ...761....9D, 2015ApJ...814L..17S}.

\begin{figure}[ht!]
\centering
\includegraphics [width=0.9\textwidth]{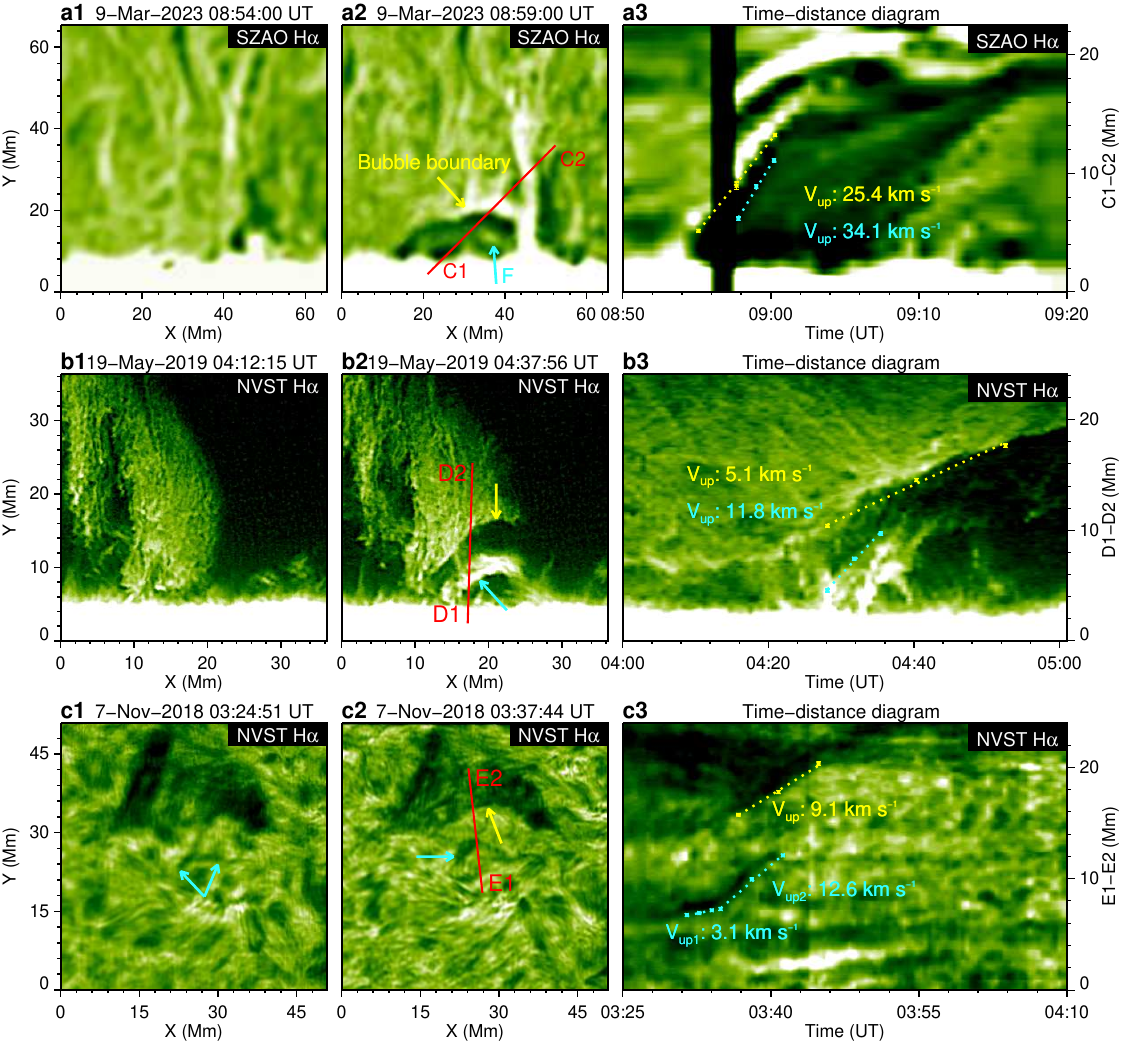}
\caption{Typical Type-II bubbles driven by erupting mini-filaments. (a1)--(a2) SZAO H$\alpha$ images displaying a host prominence without and with a Type-II bubble. The arrows mark the bubble boundary (yellow) and mini-filament (cyan). (a3) A time-distance plot derived from SZAO H$\alpha$ images along the cut ``C1--C2'' in panel (a2). The vertical black strip is a gap in observations. (b1)--(b3) and (c1)--(c3) Corresponding NVST H$\alpha$ images and time-distance plots of another two Type-II bubbles. An animation of SZAO and NVST H$\alpha$ images is available online.
}
\label{fig3}
\end{figure}

According to the observations above, we propose that it is the erupting F1 and F2 that result in the rapid formation and collapse of this Type-II bubble and subsequent plume. This conclusion is supported by the following: (1) the bubble and F1 appear on the solar limb almost at the same time; (2) the rising velocities of mini-filaments and the bubble are similar, what suggests the consistent evolution tendency; (3) the bubble collapsed after F1 and F2 reached the bubble boundary, accompanied by plume formation. It is also worth noting that there is a dark void between the bright bubble boundary and the erupting F1 during their approaching, which might share a similar scenario with the classic three-part structure of coronal mass ejections, i.e., the bright front, dark cavity, and bright core \citep{1985JGR....90..275I,2019ApJ...883...43S}. The three parts in the prominence-bubble system studied here correspond to prominence plasma pileup (the bright bubble boundary), low-density magnetic flux rope (MFR) supporting the erupting mini-filament (the dark void), and the high-density mini-filament, respectively. These observations strongly support that the formation of the Type-II prominence bubbles can be driven by erupting mini-filaments below the prominences. In addition, during the evolution process of the bubble, we find several small-scale jet-like dynamic events under the bubble, which might be closely associated with the eruption of mini-filament.

Recently, \citet{2021ApJ...923L..10C} and \citet{2022AA...659A..76W} reported a bubble with an erupting MFR below as well. However, the bubble was different from the newly-formed Type-II bubbles reported here. The erupting MFR, similiar to that in \citet{2021ApJ...923L..10C} and \citet{2022AA...659A..76W}, appeared below a Type-I bubble that has already existed and kept stable for about half an hour, and then the MFR caused expansion of the existing bubble and subsequent formation of a plume. Formation process of this Type-I bubble at its early phase was not observed. Furthermore, the possibility can not be excluded that the bubble in Figure 2 has originally been behind the solar limb. Then, the bubble could not be observed on the solar limb due to the projection effect along line-of-sight (LOS). When the mini-filaments erupted, they caused the existing bubble to expand and thus be observed above the solar limb. In this case, there is no real formation of a bubble, which is just similar to the event reported by \citet{2021ApJ...923L..10C} and \citet{2022AA...659A..76W}.

To verify our point that the Type-II bubbles are closely related to erupting mini-filaments and rule out the projection effect, we check more NVST and SZAO prominence observations and find more typical Type-II bubbles. In Figure 3, three representative events are shown, especially including an on-disk Type-II bubble, which could be employed to investigate bubbles from another perspective. Figure 3(a1) shows a host prominence without bubble. Figure 3(a2) displays a bubble formed below the prominence with a sharp arch-like boundary (marked by a yellow arrow), under which is an erupting mini-filament (marked by a cyan arrow). To study the kinematic characteristics of the bubble, we draw a time-distance plot from sequences of SZAO H$\alpha$ observations along the cut ``C1--C2'', which is in the same direction as the mini-filament eruption (Figure 3(a3)). The projective rising velocities of the bubble and the mini-filament are 25.4 km s$^{-1}$ and 34.1 km s$^{-1}$, respectively. Figures 3(b1)--(b3) and (c1)--(c3) show another two events under the prominences whose bottom edges are not shielded by the solar limb. Then, the early evolution phase, i.e., the formation process, of the bubbles could be observed clearly. The rising velocities of the second bubble and the mini-filament below are 5.1 km s$^{-1}$ and 11.8 km s$^{-1}$, respectively.

The third event is an on-disk bubble, whose morphological characteristics is similar to the on-disk bubbles reported by \citet{2021ApJ...911L...9G}. From this perspective, the slow rise and follow-up eruption of a mini-filament, as well as the subsequent formation of a nearby on-disk bubble, are unambiguously observed. The projective velocity of the mini-filament is first 3.1 km s$^{-1}$ and then 12.6 km s$^{-1}$. The induced bubble has an average velocity of 9.1 km s$^{-1}$. In addition, we check LOS magnetograms from Helioseismic and Magnetic Imager \citep[HMI;][]{2012SoPh..275..207S} on board the SDO around the on-disk bubble, but find no magnetic emergences with a similar spatial scale as the bubble, which suggests that the formation of Type-ll bubble is indeed closely associated with the eruption of mini-filament rather than magnetic emergences. The evolution characteristics of these three bubbles are similar to the bubble shown in Figure 2. And there are some voids between bubble boundaries and erupting mini-filaments as well.

\section{Discussions}

According to the presented observations of Type-II bubbles, we realize that the mini-filament eruption that appears below a quiescent prominence can form a short-lived, rapidly expanding, and quickly collapsing bubble in the dense prominence. Then, a natural question arises whether such mini-filament eruption under a prominence always leads to formation of Type-II bubble? To answer this question, we check more NVST observations of quiescent prominences and find that plumes can form either at bubble boundary after the bubble collapse or even without bubble formation. And the plumes appearing without bubbles are usually accompanied by spicule-like activities below the prominence. Similarly as \citet{2012ASPC..454...79B}, we believe that these plumes are essentially another form of Type-II bubble, but without a typical arch shape. And they are also driven by the underlying small-scale erupting activity and then directly penetrate into the prominence under certain circumstances.

\begin{figure}[ht!]
\centering
\includegraphics [width=0.9\textwidth]{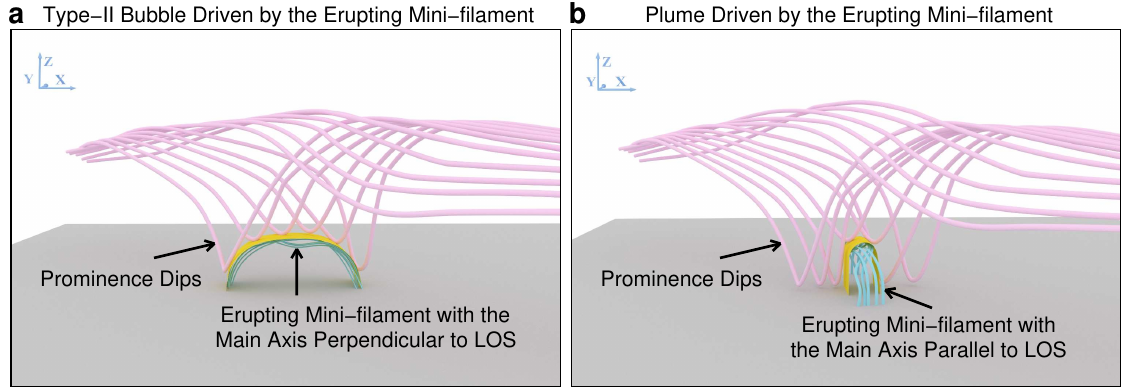}
\caption{Carton models of the Type-II bubble and plume driven by erupting mini-filaments. (a) Type-II bubble driven by erupting mini-filaments with main axis perpendicular to LOS. (b) Plume driven by erupting mini-filaments with main axis parallel to LOS. The pink curves, cyan curves, and yellow surfaces represent the prominence, the mini-filaments, and their interfaces, respectively. The black arrows point to the prominence dips and erupting mini-filaments, respectively.
}
\label{fig4}
\end{figure}

To illustrate the two scenarios when a mini-filament erupts under a prominence more specifically, we draw a carton model in Figure 4. When the angle between the axis of the erupting mini-filament (the cyan curves) and LOS direction is large enough and the erupting mini-filament can reach and penetrate into the overlying prominence (the pink curves), a Type-II bubble can be formed (Figure 4(a)). Pushed by the erupting mini-filaments, plasma piled-up in the prominence dips will form an arch-shaped high-density front (the yellow surface) with a similar scale of the mini-filament, which corresponds to the bright bubble boundary. However, if the angle is small, a narrow but large plume will be triggered by the erupting mini-filament, and the interface here corresponds to the bright boundary of the plume (Figure 4(b)). Meanwhile, due to the projection effect, the erupting filament can not be clearly detected with an arch shape and might manifest as jet or spicule-like activities in practical observations.

\section{Summary}

Based on high-resolution observations from the NVST and SZAO, we find that prominence bubbles in different events display obvious diversity of dynamic behavior. According to their dynamic properties, we firstly propose a new perspective that bubbles can be divided into two categories: quasi-steady (Type-I) bubbles and transient (Type-II) bubbles. The observational characteristics of the two types of bubbles, including the lifetime, stability duration, size, and rising velocity are summarized in Table \ref{t2}. The Type-I bubbles could remain relatively stable from 30 minutes to several hours, which can be well explained by the existing model that the interaction between the slowly-emerging or quasi-stable magnetic loops and the upper prominence dips could support the existence of prominence bubbles. On the contrary, Type-II bubbles have extremely dynamic properties of evolution and have been rarely reported previously. They usually quickly expand with a velocity of several to tens km s$^{-1}$ and collapse within one hour, below which there are erupting mini-filaments with a slightly higher rising velocity. We speculate that the visibility of Type-II bubbles is closely related to the angle between the axis of the erupting mini-filament and LOS direction: when such angle is large enough, the interaction between the erupting mini-filament and the overlying prominence could trigger a Type-II bubble with the typical arch-shaped but quickly-expending bright boundary. However, if the angle is small, a narrow but large plume will be triggered by the erupting mini-filament.

\begin{table*}
\caption{Characteristics of Type-I and Type-II Bubbles}
\centering
\begin{tabular}{cccccccc}
\tablewidth{0pt}
\hline\hline
\textbf{Bubble}&\textbf{Lifetime}&\textbf{Stability}&\textbf{Width}&\textbf{Height}&\textbf{Rising} &\textbf{Mini-filament}&\textbf{Reference}\\
& &\textbf{Duration}& & &\textbf{Velocity}&\textbf{Rising Velocity} &\\
\hline\hline

\multirow{3}*{Type-I}  &  $>$50 min--23 h  &  $\thicksim$30 min--3 h  & 12--45 Mm & 11--25 Mm & $<$3 km s$^{-1}$ & None &1--9\\
\cline{2-8}
   & \multirow{2}*{$>$14 h} &\multirow{2}*{$\thicksim$70 min} & \multirow{2}*{$\thicksim$35 Mm} & \multirow{2}*{$\thicksim$16 Mm} & \multirow{2}*{$\thicksim$0} & \multirow{2}*{None}  & The present work\\
   & & & & & & & 10 \\

\hline
\multirow{6}*{Type-II} & 30--50 min & $\thicksim$0 &  $\thicksim$25--42 Mm & None & $\thicksim$12--13 km s$^{-1}$ & $\thicksim$12--20 km s$^{-1}$ & 11\\

\cline{2-8}
 & \multirow{2}*{$\thicksim$35 min} & \multirow{2}*{$\thicksim$0} & \multirow{2}*{$\thicksim$25 Mm} & \multirow{2}*{$\thicksim$17 Mm} & \multirow{2}*{$\thicksim$7.6 km s$^{-1}$ }&$\thicksim$8.4 km s$^{-1}$  & \multirow{5}*{The present work}\\
 &  &  &  & & & $\thicksim$12.8 km s$^{-1}$  & \\

\cline{2-7}
 & $\thicksim$9 min & $\thicksim$0 & $\thicksim$16 Mm & $\thicksim$8 Mm & $\thicksim$25.4 km s$^{-1}$ & $\thicksim$34.1 km s$^{-1}$  &  \\

\cline{2-7}
 & $\thicksim$40 min & $\thicksim$0 & $\thicksim$10 Mm & $\thicksim$18 Mm & $\thicksim$5.1 km s$^{-1}$ &  $\thicksim$11.8 km s$^{-1}$   &  \\

 \cline{2-7}
 & $\thicksim$15 min & $\thicksim$0 & $\thicksim$22 Mm & $\thicksim$8 Mm & $\thicksim$9.1 km s$^{-1}$ & $\thicksim$12.6 km s$^{-1}$  &  \\

 \hline

\label{t2}
\end{tabular}
\tablecomments{
References. (1) \citet{2010ApJ...716.1288B,2011Natur.472..197B,2017ApJ...850...60B};
(2) \citet{2010SoPh..267...75R}; (3) \citet{2012ASPC..454...79B}; (4) \citet{2012ApJ...761....9D};
 (5) \citet{2012ASPC..456...77S}; (6) \citet{2014AA...567A.123G}; (7) \citet{2015ApJ...814L..17S};
 (8) \citet{2021ApJ...923L..10C}; (9) \citet{2022AA...659A..76W}; (10) \citet{2021RAA....21..222X} ; (11) \citet{2008ApJ...687L.123D}}
\end{table*}

Although bubbles and plumes have been extensively investigated in the past decade \citep{2010ApJ...716.1288B,2011Natur.472..197B,2017ApJ...850...60B,2008ApJ...687L.123D,2010SoPh..267...75R,
2012ApJ...761....9D,2015ApJ...814L..17S,2021ApJ...911L...9G,2021ApJ...923L..10C,2022AA...659A..76W,2021RAA....21..222X}, there are still many mysteries needed to be explored, especially about their formation mechanism and evolution properties. In the present work, we propose a new type of bubbles: the transient bubbles driven by erupting mini-filaments with the axis owing large angle to LOS. Moreover, we speculate that a small angle between the axis of the erupting mini-filament and LOS direction will facilitate the formation of a large plume rather than a transient bubble, which needs to be further verified by more observational evidences. In addition, there is another important topic that is not discussed in this paper: What is the physical nature of the interaction between the erupting mini-filament and the overlying prominence when the mini-filament eventually touches the prominence and drives the formation of plume and collapse of Type-II bubble? Is the obvious brightening observed at the top of the bubble boundary during this process caused by magnetic reconnections? If magnetic reconnections indeed happen, why are the two-sided-loop jets that associate with the magnetic reconnection between an erupting mini-filament and its overlying filament \citep{2019ApJ...883..104S} not observed here? Insights into these questions call for a comprehensive case study of typical events in the future.

\acknowledgments
The authors appreciate the anonymous referee for the valuable suggestions and careful corrections. Y.G and Y.H thank Prof. Xiaoli Yan for helpful discussions. The data used here are courtesy of the NVST, SZAO, SDO, and CHASE science teams. The authors are supported by the Strategic Priority Research Program of the Chinese Academy of Sciences (XDB0560000 and XDB41000000), BJAST Budding Talent Program (23CE-BGS-07), the National Natural Science Foundation of China (12273060, 12222306, 12073042, 12173083, 12003064, and 12073001), the Youth Innovation Promotion Association CAS (2023063), the National Key R\&D Program of China (2022YFF0503800 and 2019YFA0405000), the Open Research Program of Yunnan Key Laboratory of Solar Physics and Space Science (YNSPCC202211), the Yunnan Science Foundation for Distinguished Young Scholars (202101AV070004), the Yunnan Key Laboratory of Solar Physics and Space Science (202205AG070009).



\bibliography{bubbleref}{}
\bibliographystyle{aasjournal}

\end{document}